\documentclass[aps,prl,twocolumn,groupedaddress,draft,showpacs,intlimits,amsmath,amssymb,floats]{revtex4}
\usepackage{bm}
\usepackage[final]{graphicx}
\usepackage{epsfig}
\begin{document}
\title{Resonant Enhancement of Inelastic Light Scattering in Strongly
Correlated Materials}
\date{\today}
\author{A. M. Shvaika}
\email{ashv@icmp.lviv.ua}
\homepage{http://ph.icmp.lviv.ua/~ashv/}
\affiliation{Institute for Condensed Matter Physics of the National
Academy of Sciences of Ukraine, 1 Svientsitskii Street, 79011 Lviv, Ukraine}
\author{O. Vorobyov}
\email{vorobyov@icmp.lviv.ua}
\homepage{http://ph.icmp.lviv.ua/~vorobyov/}
\affiliation{Institute for Condensed Matter Physics of the National
Academy of Sciences of Ukraine, 1 Svientsitskii Street, 79011 Lviv, Ukraine}
\author{J. K. Freericks}
\email{freericks@physics.georgetown.edu}
\homepage{http://www.physics.georgetown.edu/~jkf/}
\affiliation{Department of Physics, Georgetown University, Washington, DC 20057}
\author{T. P. Devereaux}
\email{tpd@lorax.uwaterloo.ca}
\homepage{http://www.sciborg.uwaterloo.ca/~tpd/}
\affiliation{Department of Physics, University of Waterloo, Canada, Ontario
N2L 3GI}

\begin{abstract}
We use dynamical mean field theory to find an exact solution for
inelastic light scattering in strongly correlated materials such as those
near a quantum-critical metal-insulator transition. We evaluate the results for
$\textbf{q}=0$ (Raman) scattering and
find that resonant effects can be quite large, and yield a
triple resonance, a significant
enhancement of nonresonant scattering peaks, a joint resonance of both
peaks when the incident photon frequency is on the order of $U$, and the
appearance of an isosbestic point in all symmetry channels for an intermediate
range of incident photon frequencies.
\end{abstract}


\maketitle

Inelastic light scattering is a powerful tool to unravel the
nature of elementary excitations in a wide variety of
materials~\cite{review}, ranging from Kondo
insulators~\cite{Pap1,Pap2}, to high temperature
superconductors~\cite{resonance,blumberg}, to colossal
magnetoresistance materials~\cite{Pap3}. The experimental efforts
have grown tremendously with the availability of third generation
light source facilities and improvements in CCD detectors. These
efforts have been brought to bear on strongly correlated materials
to examine the elementary excitations of insulators and metals and
how they evolve as the correlations are made to change via doping, for
example.

One of the most studied areas is how the scattering cross section
resonates with the incoming light frequency. It is widely believed
that by tuning the incident photon frequency, features of the
non-resonant spectra can be magnified by orders of magnitude; that
is, the resonance serves as a bootstrap to raise the intensity of
the non-resonant signal. However, a full, consistent theory is
lacking~\cite{Platzmann,Shastry}. Non-resonant scattering is
derivable from a two-particle correlation function which can be
treated by a variety of techniques, yet the resonant and 
mixed contributions involve higher particle correlations and are
difficult to treat theoretically due to multiple-particle vertex
renormalizations. Most of the approaches to light scattering in
insulators examine the Loudon-Fleury model~\cite{LF} which is
appropriate for off-resonant conditions for the scattering of
light off of spin excitations, for example. In the strong-coupling
regime, a perturbative approach has been used to illustrate a
number of important features of electronic resonant scattering
processes~\cite{Shastry,Chubukov}.  The nonresonant case has also
been examined, and an exact solution for correlated systems (in
large spatial dimensions) is available for both the
Falicov-Kimball~\cite{paper1} and Hubbard~\cite{paper2} models.
Here we concentrate on an exact solution of the full problem for
the Falicov-Kimball model including all non-resonant, resonant and mixed
scattering channels.

For an electronic system with nearest-neighbor
hopping, the interaction with a weak external transverse
electromagnetic field $\bm A$ is described by~\cite{Shastry}
\begin{equation}\label{Hint}
    H_{\text{int}} = -\frac{e}{\hbar c} \bm j\cdot \bm A
    + \frac{e^2}{2\hbar^2 c^2}  \sum_{\alpha\beta}A_\alpha
    \gamma_{\alpha\beta} A_\beta,
\end{equation}
where
\begin{equation}\label{Hint1}
    j_\alpha = \sum_{\bm k} v_\alpha(\bm k)
    c_\sigma^\dagger({\bm k + \bm q}/2) c_\sigma({\bm k - \bm q}/2 ),
\end{equation}
is the current operator,  $v_\alpha(\bm k)=\partial\varepsilon(\bm k)/
\partial k_\alpha$ is the  Fermi velocity, and
\begin{equation}\label{Hint2}
    \gamma_{\alpha\beta} = \sum_{\bm k}
    \frac{\partial^2\varepsilon(\bm k)}{\partial k_\alpha\partial k_\beta}
    c_\sigma^\dagger({\bm k +\bm q}/2) c_\sigma({\bm k -\bm q}/2)
\end{equation}
is the stress tensor operator. The inelastic light-scattering
cross section becomes ($\Omega = \omega_i-\omega_f$, ${\bf q}={\bf
k_{i}-k_{f}}$ is the transferred photon frequency and momentum,
respectively):
\begin{equation}\label{Raman_gen}
    R(\Omega) = R_N(\Omega) + R_M(\Omega) +
                      R_R(\Omega),
\end{equation}
where the nonresonant contribution is
\begin{eqnarray}\label{Raman_N}
    R_N(\Omega) &=& 2\pi g^2(\bm k_i) g^2(\bm k_f)\nonumber
    \\
    &\times&
    \sum_{if} \frac{\exp(-\beta\varepsilon_i)}{\mathcal{Z}} \;
    \Tilde \gamma_{if} \;
    \Tilde \gamma_{fi} \;
    \delta(\varepsilon_f - \varepsilon_i - \Omega),
\end{eqnarray}
the mixed contribution is
\begin{eqnarray}\label{Raman_M}
    R_M(\Omega) &=& 2\pi g^2(\bm k_i) g^2(\bm k_f) \sum_{ifl}
\frac{\exp(-\beta\varepsilon_i)}{\mathcal{Z}}\nonumber
    \\
                &\times&
    \Biggr[
    \Tilde \gamma_{if}
    \Biggr(
    \frac{
    j^{(f)}_{fl}
    j^{(i)}_{li}
    }
    {\varepsilon_l - \varepsilon_i - \omega_i + i0^+}
    +
    \frac{
    j^{(i)}_{fl}
    j^{(f)}_{li}
    }
    {\varepsilon_l - \varepsilon_i + \omega_f - i0^+}
    \Biggr)\nonumber
    \\
    & +&
    \Biggr(
    \frac{
    j^{(i)}_{il}
    j^{(f)}_{lf}
    }
    {\varepsilon_l - \varepsilon_i - \omega_i - i0^+}
    +
    \frac{
    j^{(f)}_{il}
    j^{(i)}_{lf}
    }
    {\varepsilon_l - \varepsilon_i + \omega_f + i0^+}
    \Biggr)
    \Tilde \gamma_{fi}
    \Biggr]\nonumber
    \\
    &\times &
    \delta(\varepsilon_f - \varepsilon_i - \Omega),
\end{eqnarray}
and the resonant contribution is
\begin{eqnarray}\label{Raman_R}
    R_R(\Omega) &=& 2\pi g^2(\bm k_i) g^2(\bm k_f) \sum_{ifll'}
    \frac{\exp(-\beta\varepsilon_i)}{\mathcal{Z}}\nonumber
    \\
    &\times&\Biggr(
    \frac{
    j^{(i)}_{il}
    j^{(f)}_{lf}
    }
    {\varepsilon_l - \varepsilon_i - \omega_i - i0^+} +
    \frac{
    j^{(f)}_{il}
    j^{(i)}_{lf}
    }
    {\varepsilon_l - \varepsilon_i + \omega_f + i0^+}
    \Biggr)\nonumber
    \\
    &\times&
    \Biggr(
    \frac{
    j^{(f)}_{fl'}
    j^{(i)}_{l'i}
    }
    {\varepsilon_{l'} - \varepsilon_i - \omega_i + i0^+} +
    \frac{
    j^{(i)}_{fl'}
    j^{(f)}_{l'i}
    }
    {\varepsilon_{l'} - \varepsilon_i + \omega_f - i0^+}
    \Biggr)\nonumber
    \\
    &\times&
    \delta(\varepsilon_f - \varepsilon_i - \Omega).
\end{eqnarray}
Here $\omega_{i(f)}$ and $\bm k_{i(f)}$
denote the energy and momentum of the initial (final) states of the photons,
$\varepsilon_{i(f)}$ are the eigenvalues corresponding to the eigenstates
that describe the ``electronic matter'', and
$g(\bm k) = (hc^2/V\omega_{\bm k})^{1/2}$ is the ``scattering strength''
with  $\omega_{\bm k}=c|\bm k|$.
We have introduced the following symbols
\begin{equation}\label{Raman_not}
    \Tilde \gamma = \sum_{\alpha\beta} e_\alpha^i \gamma_{\alpha\beta}
    e_\beta^f, \quad\quad
    j^{(i),(f)} = \sum_\alpha e_\alpha^{i,f} j_\alpha,
\end{equation}
with the notation $\mathcal{O}_{if} = \left\langle i \left|
\mathcal{O} \right| f \right\rangle$ for the matrix elements of an
operator $\mathcal{O}$, $\mathcal{Z}$ the partition
function, and ${\bf e}^{i,f}$ are the incident and scattered light 
polarization vectors, respectively.  We concentrate on the
light scattering response function $\chi(\Omega)$, which is
related to the cross section, but with a Bose statistical factor
removed:
\begin{equation}
R(\Omega)=\frac{2\pi g^2(\textbf{k}_i)g^2(\textbf{k}_f)}
{1-\exp(-\beta\Omega)}\chi(\Omega).
\label{eq: chidef}
\end{equation}

Inelastic light scattering examines charge excitations of
different symmetries by employing polarizers on both the incident
and scattered light. The $A_{\textrm{1g}}$ symmetry has the full
symmetry of the lattice and is primarily measured by taking the initial and
final polarizations to be $\textbf{e}^i=\textbf{e}^f=(1,1,1,...)$.
The $B_{\textrm{1g}}$ symmetry involves crossed polarizers:
$\textbf{e}^i=(1,1,1,...)$ and $\textbf{e}^f=(-1,1,-1,1,...)$;
while the $B_{\textrm{2g}}$ symmetry is rotated by 45 degrees,
with $\textbf{e}^i=(1,0,1,0,...)$ and
$\textbf{e}^f=(0,1,0,1,...)$. While a symmetry analysis can be
employed for all momentum transfers ${\bf q}$, for Raman (${\bf
q}=0$) scattering, it is easy to show that for a system with only nearest
neighbor hopping and in the limit of large
dimensions, the $A_{\textrm{1g}}$ sector has contributions from
nonresonant, mixed, and resonant scattering, the $B_{\textrm{1g}}$
sector has contributions from nonresonant and resonant scattering
only, and the $B_{\textrm{2g}}$ sector is purely
resonant~\cite{paper1}. The symmetry analysis would be
substantially different for lower dimensions but is not currently tractable.
A full analysis for all ${\bf q}$ will be presented elsewhere and
therefore for the remainder of the paper we focus on Raman
scattering (${\bf q}=0$) only.

Normally the matrix elements defined in Eq.~(\ref{Raman_not})
cannot be easily determined for a many-body system in the
thermodynamic limit. Instead, the light scattering cross section
expressions must be evaluated by first considering the relevant
multi-time correlation functions on the imaginary time axis, then
Fourier transforming to a Matsubara frequency representation, and
finally making an analytic continuation from the imaginary to the
real frequency axis.  In the case of nonresonant scattering, the
expressions to be analytically continued depend on only one
frequency; for mixed scattering they depend on two frequencies,
and for resonant scattering, they depend on three. 
The analytic continuation procedure for the mixed and resonant Raman scattering
is complicated, because it requires a multistep procedure, where
first the transferred frequency is continued to the real axis,
then the individual initial and final frequencies are continued to
the real axis. In addition to the analytic continuation, we also
must evaluate the dressed multi-time correlation functions.  There
are renormalizations associated with two-particle ``ladder-like''
summations for a number of the relevant diagrams, but the symmetry
of the velocity operator, and of the relevant multi-particle
vertex functions (which are local in the large-dimensional limit)
imply that there are no parquet-like summations, nor are there any
three- or four-particle vertex renormalizations~\cite{footnote}.
Since the two-particle vertex function for the Falicov-Kimball
model is already known~\cite{SFM}, the full Raman scattering
problem can be solved via a straightforward but tedious procedure.
The final formulas are cumbersome and will be presented elsewhere.

The Falicov-Kimball model Hamiltonian satisfies~\cite{FK}
\begin{equation}
H=-\frac{t^*}{2\sqrt{d}}\sum_{\langle
ij\rangle}(c^\dagger_ic_j+c^\dagger_jc_i)
+U\sum_ic^\dagger_ic_iw_i \label{eq: hamiltonian}
\end{equation}
where $c^\dagger_i$ ($c_i$) create (annihilate) a conduction
electron at site $i$, $w_i$ is a classical variable (representing
the localized electron number at site $i$) that equals 0 or 1,
$t^*$ is a renormalized hopping matrix that is nonzero between
nearest neighbors on a hypercubic lattice in $d$-dimensions (and
we take the limit $d\rightarrow\infty$~\cite{metzner_vollhardt}),
and $U$ is the local screened Coulomb interaction between
conduction and localized electrons. This model can be solved
exactly by using dynamical mean field theory, as described by
Brandt and Mielsch~\cite{BM} and summarized in review
articles~\cite{reviews}.

We concentrate on the case with $U=2$ here, which is just on the
insulating side of the metal-insulator transition at half filling
($\rho_e=\langle w_i \rangle =1/2$). This was the regime where the
nonresonant response showed a number of interesting properties for
both Raman~\cite{paper1} and inelastic x-ray
scattering~\cite{paper3}.  The Stokes Raman response function is
plotted in Fig.~\ref{fig: raman_full} at $T=0.5$ for 9 different
incident photon frequencies $\omega_{i}$
ranging from 0.5 to 4.5 in steps of 0.5.
Since the transferred energy can be no larger than the
incident photon energy, all scattering curves run from zero up to
$\omega_i$.  The first thing to note in Fig.~\ref{fig: raman_full}
is the large nearly vertical line that occurs as
$\Omega\rightarrow\omega_i$.  This is the triple
resonance~\cite{Chubukov}, which yields a strong enhancement to
the Raman scattering when the energy of the scattered photon
approaches zero.
In the Loudon-Fleury regime~\cite{LF}, where the
incident photon energy is much larger than the electronic
excitation energies, we see that the full response is essentially
that of the nonresonant response~\cite{paper1} plus the
triple-resonance peak. As the incident photon energy is reduced,
the behavior becomes much more complex. Generically we can
identify a number of low-energy and higher-energy resonant
enhancements to the scattering.

\begin{figure}[htbf]
\epsfxsize=3.0in \epsffile{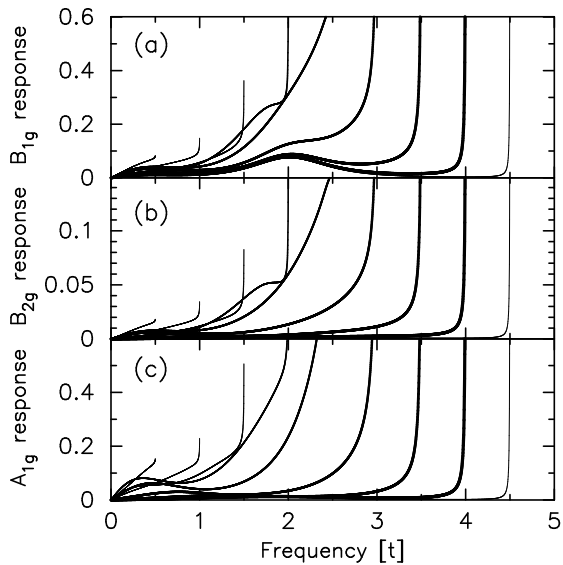} \caption{\label{fig:
raman_full} Raman response function for different channels.  We
take $U=2$, $T=0.5$, and choose $\omega_i=0.5-4.5$ in steps of 0.5
(the different line thicknesses correspond to different
$\omega_i$'s).
}
\end{figure}



\begin{figure}[htbf]
\epsfxsize=3.0in
\epsffile{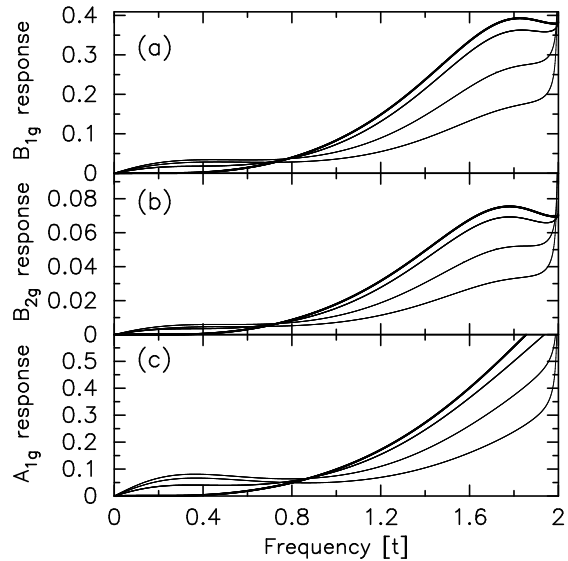}
\caption{\label{fig: raman_iso}
Raman response function for $U=2$ and $\Omega=2$ for
different channels at $T=1$, $T=0.5$, $T=0.2$,
and $T=0.05$.  The temperature decreases as the lines are made thicker.
}
\end{figure}

One interesting feature of the response function, seen in
experiments on correlated materials~\cite{Pap1,Pap2}, and seen in
theoretical calculations of the nonresonant
response~\cite{paper1,paper2}, is that at low energy there is an
isosbestic point, where the $B_{\textrm{1g}}$ response function is
essentially independent of temperature at a particular frequency
$\Omega\approx U/2$.  Below that frequency the response decreases
as $T$ is lowered, and above it increases. The isosbestic behavior
must survive in the Loudon-Fleury regime, because the isosbestic
point is at low energy, and the low-energy response is negligible
in the resonant and mixed contributions.  But what happens when
$\omega_i\approx U$?  Here we expect interesting effects to occur,
because the incident photon energy is the right size to cause
transitions from the lower to upper Hubbard bands of the
correlated insulator.  Indeed, we find interesting results in this
regime (Fig.~\ref{fig: raman_iso}).  At low temperature
$(T<0.7)$, a symmetry-dependent isosbestic point appears at a
transferred frequency of 0.7--0.9 and is seen in all channels at
low enough $T$, even the $A_{\textrm{1g}}$ and $B_{2g}$ channels, which 
have no isosbestic point in the nonresonant regime.  Hence the inclusion
of resonant and mixed terms provides theoretical support for the
generic presence of a low-temperature isosbestic point in
correlated systems.

\begin{figure}[htbf]
\epsfxsize=3.0in
\epsffile{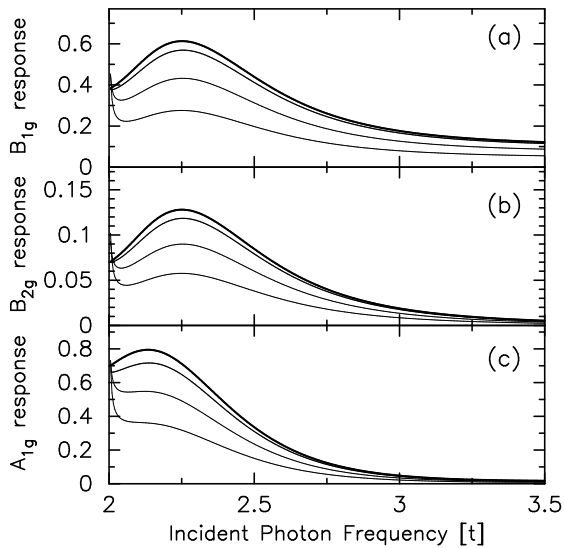}
\caption{\label{fig: raman_prof1}
Raman response function for $U=2$ and $\Omega=2$ for
different channels at $T=1$, $T=0.5$, $T=0.2$,
and $T=0.05$.  The temperature decreases as the lines are made thicker.
}
\end{figure}

\begin{figure}[htbf]
\epsfxsize=3.2in
\epsffile{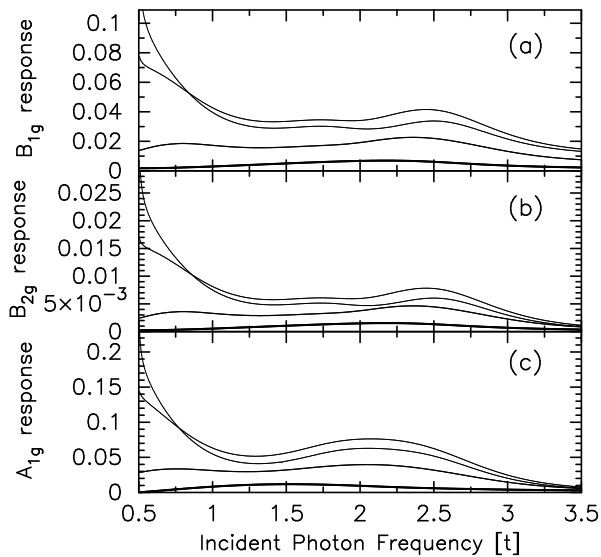}
\caption{\label{fig: raman_prof2}
Raman response function for $U=2$ and $\Omega=0.5$
for different channels at $T=1$, $T=0.5$, $T=0.2$,
and $T=0.05$.  The temperature decreases as the lines are made thicker.
}
\end{figure}

Finally, we present results of what the resonant profile of the
scattering looks like by fixing the transferred frequency and
varying the incident photon energy.  We expect that there will be
a resonant peak in the response, and indeed this is so, although
in some cases the triple resonance overwhelms the presence of the
peak. Note that in high-temperature superconductors, in addition
to the expected resonance that occurs when the incident photon
frequency is close to the transferred frequency, another resonance
occurs, where the low-energy peak is strongly enhanced when
$\omega_i\approx U$~\cite{resonance,blumberg}.  We show this
regime in Figs.~\ref{fig: raman_prof1} ($\Omega=2.0$) and
\ref{fig: raman_prof2} ($\Omega=0.5$). In Fig.~\ref{fig:
raman_prof1} we see a moderately broad peak centered at $\omega_i$
10--20\% higher than $U$.  The enhancement of the charge-transfer
peak in this regime can easily be an order of magnitude over the
nonresonant response. In Fig.~\ref{fig: raman_prof2}, we see a
similar resonant feature when the incident photon frequency is
slightly larger than $\Omega =0.5$ (arising from the
triple-resonance effect), but a second less prominent series of
broad peaks occurs when $\omega_i\approx U$ indicating that the
low-energy and charge transfer peaks are resonating together when
$\omega_i\approx U$.  Hence the behavior observed in the
high-temperature superconductors~\cite{resonance,blumberg} is
likely to be seen in many other correlated insulators.

In conclusion, we have shown a number of interesting resonant features
in theoretical calculations of electronic Raman scattering.
These features include the triple resonance, the resonant enhancement of
nonresonant peaks, the appearance of isosbestic points, and the joint
resonance of low-energy
and charge-transfer peaks when $\omega_i\approx U$. It will
be interesting to see whether these predictions can be seen in future
experiments on correlated systems.

\acknowledgments
We wish to acknowledge the U. S. Civilian Research and Development Foundation 
through grant number UP2-2436-LV-02.  J.~K.~F.~ also acknowledges the National
Science Foundation through grant number DMR-0210717, and T.~P.~D. acknowledges
funding from NSERC, PREA, and the Alexander von Humboldt Foundation.


\begin{thebibliography}{99}
\bibitem{footnote} We have not been able to prove that the three and
four-particle vertices do not enter for the resonant Raman
scattering, but a strong-coupling perturbative analysis indicates
that this is so to lowest order.
\end{thebibliography}
\end{document}